# A Study of Cosmic Expansion Generated by Non-conservation of Matter in the Framework of Brans-Dicke Theory

Sudipto Roy[1] and Mohsin Islam[2]

[1]Department of Physics,

[2]Department of Mathematics & Statistics (Commerce Evening),

St. Xavier's College, 30 Mother Teresa Sarani (Park Street),

Kolkata 700016, West Bengal, India.

Emails: (1) roy.sudipto1@gmail.com, (2) mislam416@gmail.com

**Abstract:** The present study, on the expansion of universe, is based on an assumption regarding the possibility of inter-conversion between matter and dark energy, through some interaction of matter with the scalar field in the framework of Brans-Dicke theory. The field equations for a spatially flat space-time have been solved using an empirical dependence of scalar field parameter upon the scale factor. To represent the behaviour regarding the non-conservation of matter, a function, expressed in terms of the Hubble parameter, has been empirically incorporated into the field equations. Their solution shows that, this function, whose value is proportional to the matter content of the universe, decreases monotonically with time. This matter-field interaction generates late time acceleration, causing the deceleration parameter to change its sign from positive to negative. Time dependence of the proportion of dark energy component of the universe has been determined and shown graphically. Time variation of gravitational constant and the Brans-Dicke dimensionless parameter has been analyzed in the present study. The rate of generation of dark energy from matter has been found to affect the time variations of deceleration parameter and gravitational constant.

## 1. Introduction

On the basis of recent data obtained from the Wilkinson Microwave Anisotropy Probe (WMAP) and observational information from the Supernova Cosmology Project, it has been strongly confirmed that the universe is expanding with acceleration [1-8]. According to these observations, nearly 70% of the total energy density of the universe, apparently unclustered, has a large negative pressure and is referred to as dark energy (DE), and it is widely regarded as responsible for accelerated cosmic acceleration. The search for this new entity, which is believed to cause this acceleration, has not seen a preferred direction as yet. The cosmological constant ($\lambda$), a well known parameter of the general theory of relativity, is one of the most suitable candidates acting as the source for this repulsive gravitational effect and it fits the observational data reasonably well, but it has its own limitations [9, 10]. A large number of possible candidates for this dark energy component has already been proposed and their behaviour have been studied extensively [9, 11]. It is worth mentioning that this accelerated expansion is a very recent phenomenon and it follows a phase of decelerated expansion. This is important for the successful nucleosynthesis and also for the structure formation of the universe. As per observational findings, beyond a certain value of the redshift (z) (i.e. $z > 1.5$ ), the universe surely had a decelerated phase of expansion [12]. So, the dark energy component should have evolved in such a way that its effect on the dynamics of the universe is dominant only during later stages of the matter dominated era. A recent study by Padmanabhan and Roy Choudhury [13], through analysis of the supernova data, shows that the deceleration parameter ($q$) of the universe has surely undergone a signature flip from positive to negative.

Apart from the cosmological constant ($\lambda$), a large number of other models of dark energy have already appeared in the literature with their own characteristic features [15-17]. All these models have been found to generate the cosmic acceleration very effectively. Out of these candidates, one of the most popular candidates is a scalar field with a positive potential which can generate an effective negative pressure if the potential term dominates over the kinetic term. This scalar field is often called the *quintessence scalar field*. A large number of quintessence potentials have appeared in scientific literature and their behaviours have been studied extensively. One may go through a detailed study by V Sahni in this regard [18]. However, there is no proper physical background or interpretation of the origin of most of the quintessence potentials.

A method to avoid ambiguities in these models is to consider that the two components of matter, namely, the CDM and dark energy or the Q-matter are interacting so that during this interaction there will be some transfer of energy from one field to another. A number of models have been proposed where a transfer of energy takes place from the component of dark matter to the component of dark energy [19-25], so that during the late time of evolution, the dark energy predominates over the ordinary matter and causes acceleration of the universe. However, most of these models are found to be based on interactions that have been chosen arbitrarily, without being supported by any physical theory. There has been a prolonged search for a suitable cosmologically viable interacting model of dark energy.

To avoid ambiguities and difficulties caused by the arbitrariness in the formulation of a particular Q-field, non-minimally coupled scalar field theories have shown much greater effectiveness in carrying out the possible role of an agent responsible for the late time acceleration. This is simply due to the presence of the scalar field in the purview of the theory and does not have to be introduced separately. The Brans-Dicke (BD) theory is considered to be the most natural choice as the scalar-tensor generalization of general relativity (GR), due to its simplicity and a possible reduction to GR in some limit. For this reason, Brans–Dicke theory or its modified versions have been shown to generate the present cosmic acceleration [26, 27]. It has also been shown that BD theory can potentially generate sufficient acceleration in the matter dominated era even without any help from an exotic Q - field [28]. But one actually needs a theory which can account for a transition from a state of deceleration to acceleration. Amongst other non-minimally coupled theories, a dilatonic scalar field was also considered to play a role in driving the present acceleration [29]. The dark energy and dark matter components are considered to be non-interacting in most of the models and are allowed to evolve independently. Since these two components are of unknown nature, the interaction between them is expected to provide a relatively generalized framework for study. Zimdahl and Pavon [30] have recently shown that the interaction between dark energy and dark matter can be very useful in solving the coincidence problem ( see also ref [31] ). Following this idea, one may consider an interaction or inter-conversion of energy between the Brans - Dicke scalar field which is a geometrical field and the dark matter. The concept of an inter-conversion of energy between matter and the non-minimally coupled field had been used earlier by Amendola [32].

It has been found that in most of the models in the Brans-Dicke framework, the accelerated expansion of the universe requires a very low value of $\omega$, typically of the order of unity. However, a recent work shows that if the BD scalar field interacts with the dark matter, a generalized BD theory can perhaps serve the purpose of driving acceleration even with a high value of $\omega$ [33]. In all these studies, either Brans-Dicke theory is modified to meet the present requirement or, one chooses a quintessence scalar field to generate the required acceleration. In a recent study by Barrow and Clifton[34] and also in reference [33], no additional potential was required, but an interaction between the BD scalar field and the dark matter was used to account for the observational facts.

In the present work, we have used a generalized form of Brans-Dicke theory proposed by Bergman and Wagoner [35] and expressed in a more useful form by Nordtvedt [36]. In this generalization, the parameter $\omega$ is taken to be a function of the BD scalar field $\varphi$ instead of being treated as a constant. Different functional forms of $\omega(\varphi)$ can be formulated on the basis of various physical motivations.

In the present study we have not considered any particular mechanism of interaction between matter and the scalar field. Our model is mainly based on an assumption of non-conservation of matter, keeping the possibilities open for an inter-conversion between matter and some other form of energy, may be dark energy which is widely held responsible for accelerated expansion of universe. We have introduced a function $f(t)$in the relation regarding the density of matter ($\rho$), expressed by equation (5), to account for the non-conservation of matter. It is quite evident from this equation that if we consider $f(t) = 1$ at all time, the equation expresses conservation of the matter content of the universe. An empirical form of $f(t)$ has been assumed in terms of Hubble parameter under the restriction that $f(t) = 1$ at the present instant ($t = t_0$), to have consistency with the form of equation (5). Solution of the differential equation, involving this empirical form of $f(t)$, leads to a form of deceleration parameter which changes its sign from positive to negative, as time goes on, implying a transition of the universe from a phase of deceleration to acceleration. This behaviour establishes the validity of our assumption regarding $f(t)$. The purpose of this study is to determine the nature of variation of this function $f(t)$, which can be shown to be equal to the ratio of matter content of universe at any time ($t$) to the matter content at the present time ($t_0$). The present model shows this function to be decreasing with time, at a gradually decreasing rate, implying a conversion of matter into dark energy, with a rate that slows down with time. This model thus shows that an assumption of non-conservation of matter leads to the correct behaviour of the deceleration parameter and also generates correct values of $H_0$ and $q_0$ at $t = t_0$ by a proper tuning of values of the constants. Therefore we expect to get a correct time dependence of matter density ($\rho$) from this model and we have compared this with the density obtained from an assumption of conservation of matter. Here we have explored the time dependence of the production of dark energy from matter. We have also studied here the time dependence of gravitational constant ($G$) and $\omega$ and obtained results consistent with other studies [33, 37-40].

## 2. Solution of Field equations

The field equations in the generalized Brans-Dicke theory, for a spatially flat Robertson-Walker space-time, are given by [37],

$$3\left(\frac{\dot{a}}{a}\right)^2 = \frac{\rho}{\varphi} + \frac{\omega(\varphi)}{2}\left(\frac{\dot{\varphi}}{\varphi}\right)^2 - 3\frac{\dot{a}}{a}\frac{\dot{\varphi}}{\varphi}, \tag{1}$$

$$2\frac{\ddot{a}}{a} + \left(\frac{\dot{a}}{a}\right)^2 = -\frac{\omega(\varphi)}{2}\left(\frac{\dot{\varphi}}{\varphi}\right)^2 - 2\frac{\dot{a}}{a}\frac{\dot{\varphi}}{\varphi} - \frac{\ddot{\varphi}}{\varphi}. \tag{2}$$

Combining (1) and (2) one gets,

$$2\frac{\ddot{a}}{a}+4\left(\frac{\dot{a}}{a}\right)^2=\frac{\rho}{\varphi}-5\frac{\dot{a}}{a}\frac{\dot{\varphi}}{\varphi}-\frac{\ddot{\varphi}}{\varphi}\,. \tag{3}$$

In many studies, the content of matter (dark + baryonic) of the universe is assumed to be conserved [37]. In those cases, this conservation is mathematically expressed as,

$$\rho a^3 = \rho_0 a_0^3 \tag{4}$$

Here $a_0$ and $\rho_0$ are the scale factor and the matter density of the universe respectively at the present time. According to some studies, the matter content of the universe may not remain proportional to $\rho_0 a_0^3$ [33, 38, 41]. There are some studies on Brans-Dicke theory of cosmology where one takes into account an interaction between matter and the scalar field. A possibility of an inter-conversion between these two entities remains open in these studies. Keeping in mind this possibility, we propose the following relation for the density of matter (dark matter + baryonic matter).

$$\rho a^3 = f(t)\,\rho_0 a_0^3 \tag{5}$$

In the present study we have not considered any theoretical model to explain or analyse the interaction mechanism between matter and scalar field. We have only considered the results of different such studies where it is clearly shown that equation (4) is not valid for a process through which we may have an inter-conversion between matter and scalar field. Considering non-conservation of matter, we have incorporated a function of time $f(t)$ in the relation (4) to get a new relation represented by equation (5). Thus the function $f(t)=\frac{\rho a^3}{\rho_0 a_0^3}$, is the ratio of matter content of the universe at any instant of time $t$, to the matter content at the present instant ($t=t_0$). Since the denominator is of this ratio is a constant, $f(t)$ can be regarded as a measure of the total content of matter (dark+baryonic) of the universe at the instant of time $t$. In the present study we have denoted this ratio by $R_1$ where $R_1 = M(t)/M(t_0)$. We have defined a second ratio $R_2=\frac{1}{f}\frac{df}{dt}=\frac{1}{M}\frac{dM}{dt}$ which represents fractional change of matter per unit time. If, at any instant, $R_2$ is negative, it indicates a loss of matter or a change of matter into some other form due to its interaction with the scalar field. We have also defined a third ratio $R_3=f-1=\frac{M(t)-M(t_0)}{M(t_0)}$ indicating a fraction change of matter content from its value at the present time.

The form of equation (5) makes it necessary that $f(t)=1$ at $t=t_0$ where $t_0$ denotes the present instant of time when the scale factor $a=a_0=1$ and the density $\rho=\rho_0$.

To make the differential equation (3) tractable, let us propose the following ansatz.

$$\varphi=\varphi_0(a/a_0)^{-3} \tag{6}$$

Here $\varphi$ has been so chosen that its nature of dependence upon scale factor ($a$) is the same as the nature of dependence of the matter density ($\rho$) upon the scale factor.

Combining the equations (5) and (6) with equation (3) we get,

$$\frac{\ddot{a}}{a}-\left(\frac{\dot{a}}{a}\right)^2=-f\frac{\rho_0}{\varphi_0} \tag{7}$$

In terms of Hubble parameter $H = \frac{\dot{a}}{a}$, equation (7) takes the following form.

$$\left(\dot{H} + H^2\right) - H^2 = \frac{dH}{dt} = -f\frac{\rho_0}{\varphi_0} \tag{8}$$

For equation (8), we propose the following empirical expression of the function $f(t)$.

$$f(t) = Exp[n(H - H_0)] \tag{9}$$

This choice of $f(t)$ ensures that $f(t) = 1$ at $t = t_0$. Here $n$ is a constant parameter and $H_0$ is the value of the Hubble parameter at the present instant $t_0$.
Substituting for $f(t)$ in equation (8) from equation (9) we get,

$$\frac{dH}{dt} = -\frac{\rho_0}{\varphi_0}\ Exp[n(H - H_0)] \tag{10}$$

Using $H = \dot{a}/a$ and solving the differential equation (10) we have the following expression of the scale factor.

$$a = Exp\left[\frac{t}{n}\{1 - ln(-l_2 + l_1 t)\} + \frac{l_2 ln(l_2 - l_1 t)}{n l_1} - \frac{C_2}{n}\right] \tag{11}$$

Here, $l_1 = n\frac{\rho_0}{\varphi_0}Exp(-nH_0)$ and $l_2 = nC_1 Exp(-nH_0)$. where $C_1$ and $C_2$ are the constants of integration.
Using equation (11), the Hubble parameter $(H)$ and deceleration parameter $(q)$ are obtained as,

$$H = \frac{1}{a}\frac{da}{dt} = -\frac{ln(-l_2 + l_1 t)}{n} \tag{12}$$

$$q = -\frac{\ddot{a}\,a}{\dot{a}^2} = -1 + \frac{n l_1}{(-l_2 + l_1 t) \times [ln(-l_2 + l_1 t)]^2} \tag{13}$$

Equation (11) can be written as,

$$a = Exp\left[\frac{t}{n}\{1 - ln(x)\} + \frac{l_2 ln(-x)}{n l_1} - \frac{C_2}{n}\right] \tag{14}$$

Where $x = -l_2 + l_1 t$ is a function of time.
Equation (14) contains $ln(x)$ and $ln(-x)$ where $x$ is a function of time. One of these two functions will always produce an imaginary result. It would be physically unacceptable to have the logarithm of a negative number as a part of the expression of scale factor $(a)$ which is real. To avoid this discrepancy, we take $C_1 = 0$ leading to $l_2 = 0$. Therefore, the expression of the scale factor takes the following form.

$$a = Exp\left[\frac{t}{n}\{1 - ln(l_1 t)\} - \frac{C_2}{n}\right] \tag{15}$$

Thus, $a_0 = Exp\left[\frac{t_0}{n}\{1 - ln(l_1 t_0)\} - \frac{C_2}{n}\right]$

With $l_2 = 0$, the expressions of Hubble parameter takes the following form.

$$H = \frac{1}{a}\frac{da}{dt} = -\frac{ln(l_1 t)}{n} \tag{16}$$

Using this time dependence of $H$, we may write a new expression of this parameter to ensure that $H = H_0$ at $t = t_0$. This new expression may be written as,

$$H = H_0 \frac{-\frac{ln(l_1 t)}{n}}{\left[-\frac{ln(l_1 t)}{n}\right]_{t=t_0}} = H_0 \frac{ln(l_1 t)}{ln(l_1 t_0)} \tag{16A}$$

The deceleration parameter in the present model is,

$$q = -\frac{\ddot{a}\, a}{\dot{a}^2} = -1 + \frac{n/t}{[ln(l_1 t)]^2} \tag{17}$$

Using equation (16), we may write equation (17) as,

$$q = -\frac{\ddot{a}\, a}{\dot{a}^2} = -1 + \frac{1}{n\, t\, H^2} \tag{17A}$$

Using the fact that $q = q_0$ at $t = t_0$, equation (17A) yields the following value of the parameter $n$.

$$n = \frac{1}{t_0 {H_0}^2}\frac{1}{1+q_0} \tag{18}$$

Substituting for the values of constants in equation (18) we get,
$n = 9.271 \times 10^{17}$
Combining equation (6) with (2) and using $q = -\ddot{a}a/\dot{a}^2$ we get the following expression for the dimensionless Brans-Dicke parameter $\omega$.

$$\omega = -\frac{2}{9}(7 + q) \tag{19}$$

The gravitational constant, which is known to be the reciprocal of the Brans-Dicke scalar field parameter ($\varphi$) is given by,

$$G = \frac{1}{\varphi} = \frac{(a/a_0)^3}{\varphi_0} \tag{20}$$

An experimentally measurable parameter $\frac{\dot{G}}{G}$ is given by,

$$\frac{\dot{G}}{G} = \frac{1}{G}\frac{dG}{dt} = 3\frac{\dot{a}}{a} = 3H \tag{21}$$

Using equation (21) we get,

$$\left(\frac{\dot{G}}{G}\right)_{t=t_0} = 3H_0 = 2.204 \times 10^{-10}\, Yr^{-1} \tag{22}$$

Combining equation (16A) with (9) we get,

$$R_1 = f = Exp\left[nH_0\left\{\frac{ln(l_1 t)}{ln(l_1 t_0)} - 1\right\}\right] \tag{23}$$

Using equation (23) one gets,

$$R_2 = \frac{1}{f}\frac{df}{dt} = \frac{nH_0}{ln(l_1 t_0)}\ \frac{1}{t} = (-4.58 \times 10^{-1})\ \frac{1}{t} \tag{24}$$

$$R_3 = f - 1 = Exp\left[nH_0\left\{\frac{ln(l_1 t)}{ln(l_1 t_0)} - 1\right\}\right] - 1 \tag{25}$$

Equation (24) shows that $R_2$ has a negative value which clearly indicates that the matter content of the universe decreases with time. It may be caused due to its conversion into other forms, may be dark energy. One may assume that the process of conversion of matter into dark energy started in the past at the time of $t = \alpha t_0$ where $\alpha < 1$. Hence, the value of $R_1$ at $t = \alpha t_0$ was proportional to the total matter content when no dark energy existed. To measure the proportion of dark energy in the universe, assuming matter to be its only source, we define a quantity $R_4$ in the following way.

$$R_4 = \frac{R_1(\alpha t_0) - R_1(t)}{R_1(\alpha t_0)} = \frac{M(\alpha t_0) - M(t)}{M(\alpha t_0)},\ \alpha < 1 \tag{26}$$

If the matter content of universe is assumed to be conserved, equation (8) takes the following form.

$$\frac{dH}{dt} = -\ \frac{\rho_0}{\varphi_0} \tag{27}$$

Solution of equation (27) is expressed as,

$$a = a_0 Exp\left[-\frac{\rho_0}{2\varphi_0}(t - t_0)^2 + H_0(t - t_0)\right] \tag{28}$$

To derive this scale factor we have used the conditions that,

i) $a = a_0$ and ii) $H = H_0$ at $t = t_0$.

Equation (28) leads to the following expression of density.

$$\rho = \rho_0\left(\frac{a_0}{a}\right)^3 = \rho_0\ Exp\left[\frac{3\rho_0}{2\varphi_0}(t - t_0)^2 - 3H_0(t - t_0)\right] \tag{29}$$

The expression of deceleration parameter obtained from equation (28) is,

$$q = -1 + \frac{\frac{\rho_0}{\varphi_0}}{\left[H_0 - \left(\frac{\rho_0}{\varphi_0}\right)(t - t_0)\right]^2} \tag{30}$$

## 3. Graphical analysis of theoretical findings

Figure 1 shows two plots of the deceleration parameter, as functions of time. The dashed curve is based on equation (30), derived on the assumption of conservation of matter. Its negative value indicates accelerated expansion of universe right from the beginning. The solid curve is based on equation (17A), derived on an assumption of inter-conversion of matter and energy. This curve changes sign from positive to negative indicating a transition from a phase of deceleration to acceleration of cosmic expansion.

In figure 2, we show two plots of matter density ($\rho$) of the universe, as functions of time. The dashed curve is based on equation (29), derived on the assumption of conservation of matter. The solid curve is based on equation (5), considering an inter-conversion of matter and energy. In the latter case the density decreases at a faster rate.

Figure 3 shows a plot of $R_1(\equiv f)$ as a function of time. This parameter is proportional to the matter content of the universe and it is found to decrease with time.

Figure 4 shows the variation of $R_2$ as a function of time. Its negative value clearly indicates a reduction of matter content with time, implying a conversion of matter into dark energy. Its value decreases at a gradually slower rate, indicating a gradual decrease in rate of matter-energy conversion. Its value becomes asymptotic to zero with time.

In figure 5, we show a plot of $R_3$ as a function of time. Its value is positive in the past, zero at the present time and negative in the future, as expected from the plot of $R_1$. This curve also indicates a conversion of matter into energy, at a gradually slower rate.

A plot of $R_4$ as a function of time, for three different values of the parameter $\alpha$, has been shown in figure 6. This plot is based on equation (26). It shows a relative measure of dark energy with respect to the total matter-energy content of the universe. For $\alpha = 0.07$ we find that the dark energy percentage (i.e. $100 \times R_4$) is around seventy at the present time ($t = t_0$), which is close to values obtained from other studies [33].

We have shown the variation of $G/G_0$ as a function of time in figure (7). It increases with time, as also found in other recent studies [39]. The dashed curve shows the variation under an assumption of conservation of matter, represented by equation (4). The solid curve corresponds to the situation where matter is converted into dark energy. Gravitational constant increases more rapidly in the latter case.

Figure 8 shows a plot of $\boldsymbol{\dot{G}/G}$ as a function of time. This is fractional change of gravitational constant per year. Its value is positive and it decreases with time. The value of this quantity at the present time ($\boldsymbol{t = t_0}$), according to equation (22), is less than its upper limit $\boldsymbol{4 \times 10^{-10}\, Yr^{-1}}$, as specified by Weinberg [40]. The dashed curve shows the variation of $\boldsymbol{\dot{G}/G}$ under an assumption of conservation of matter, represented by equation (4). The solid curve corresponds to the situation where matter is converted into dark energy. The dashed curve shows a much slower decrease compared to the solid curve.

The values of different cosmological parameters used in the present study are,

$H_0 = 72\left(\frac{Km}{Sec}\right) per\ Mega\ Parsec\ = 2.33 \times 10^{-18} sec^{-1}$

$t_0 = 14\ billion\ years = 4.415 \times 10^{17} sec$

$\varphi_0 = \frac{1}{G_0} = 1.498 \times 10^{10} m^{-3} Kg s^2$

$\rho_0 = 2.831 \times 10^{-27} Kg/m^3$ [present density of matter (dark+baryonic)]

$q_0 = -0.55$

## 4. Conclusions

In the present study it has been found that a theoretical possibility of an inter-conversion between matter and energy of the universe generates late time acceleration smoothly. The form of the function $\boldsymbol{f(t)}$, chosen in this model to account for the non-conservation of energy, may be varied to get a better result. It is found that this function, which is proportional to the matter content of the universe, decreases monotonically with time, indicating clearly a one-way conversion from matter component to the component dark energy. It is evident from the graphs that, as the proportion of dark energy content ($\boldsymbol{R_4}$) increases with time, the deceleration parameter gradually makes a transition towards negative values from positive ones, the gravitational constant also increases. The graphs also show that, as the rate of production of dark energy decreases, the rate of change of deceleration parameter and gravitational constant also decreases. These observations lead to an inference regarding the possibility of a dependence of the gravitational constant and the deceleration parameter on the content of dark energy and the rate at which it changes. The present model also shows that if the dark energy is assumed to have been generated entirely from matter, its present proportion depends on the time at which this process of conversion started. The earlier it began, the greater the percentage of dark energy present in the universe and it is evident from the role of the parameter $\boldsymbol{\alpha}$ in the expression of $\boldsymbol{R_4}$. One has ample scope to improve this model by changing the ansatz regarding the Brans-Dicke scalar field parameter ($\boldsymbol{\varphi}$) and also by changing the dependence of $\boldsymbol{f(t)}$ upon the Hubble parameter. Several forms of $\boldsymbol{f(t)}$ can be chosen, satisfying the simple requirement that that $\boldsymbol{f(t) = 1}$ at $\boldsymbol{t = t_0}$. It would be interesting to find out whether all of them come out to be monotonically decreasing functions of time, implying clearly a single direction of conversion, that is from matter to dark energy.

# Figures

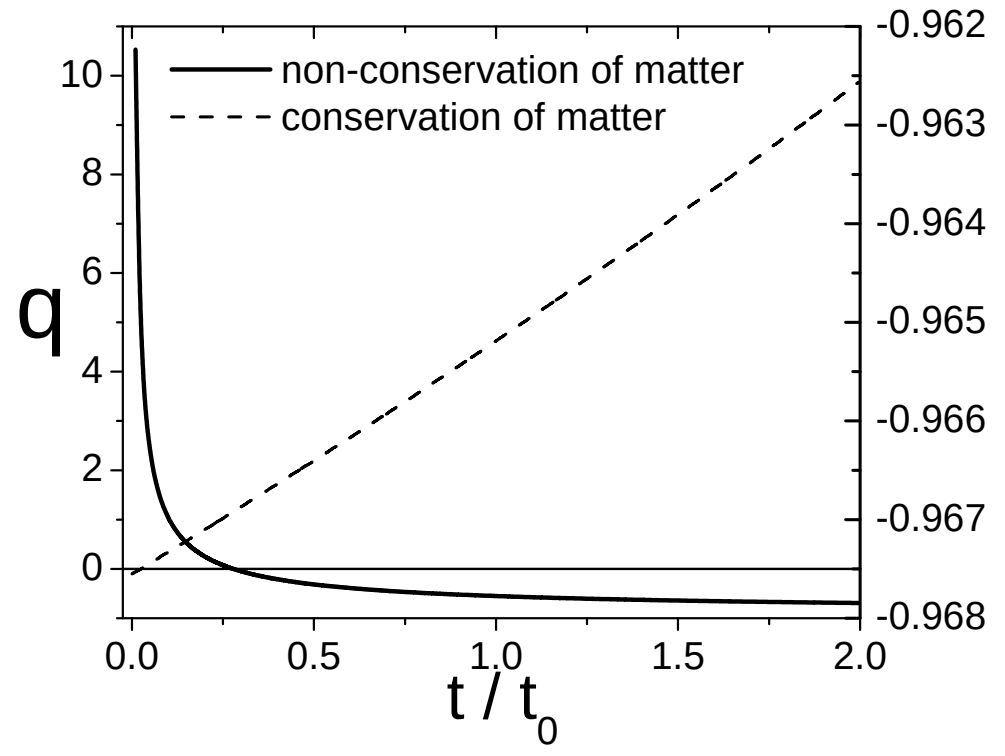

Figure 1: Plot of deceleration parameter as a function of time. The right vertical axis shows the scale for the dashed curve.

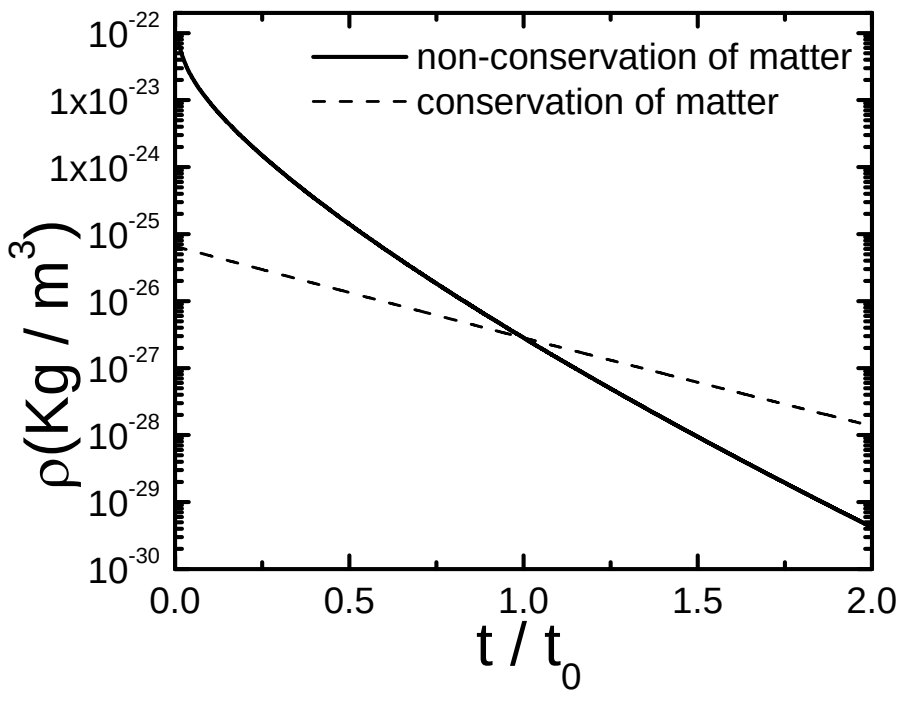

Figure 2: Plot of density of matter of the universe as a function of time.

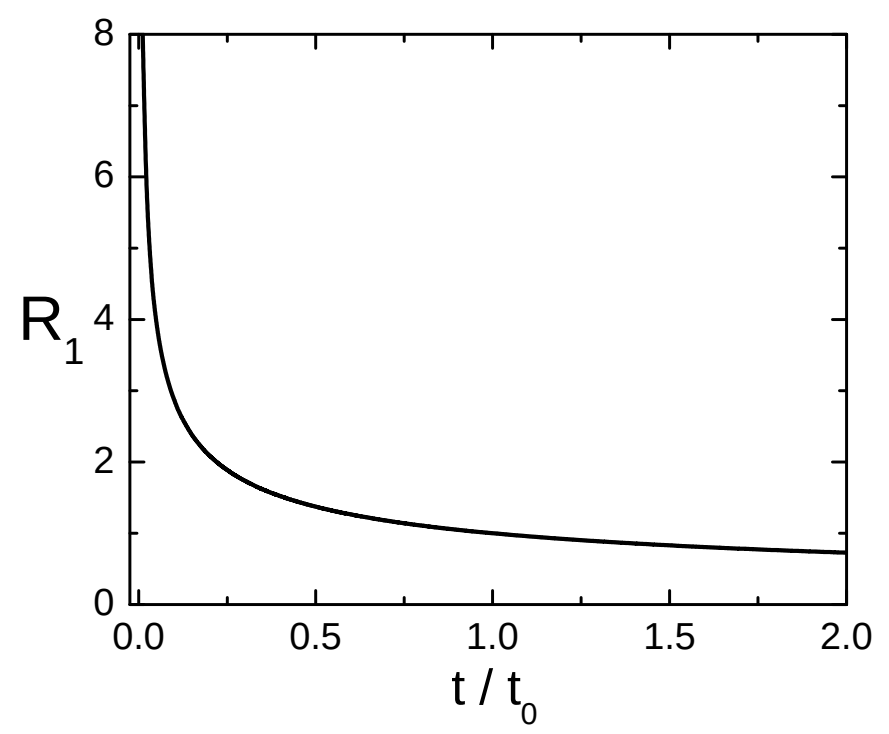

Figure 3: Plot of $R_1(\equiv f)$ as a function of time.

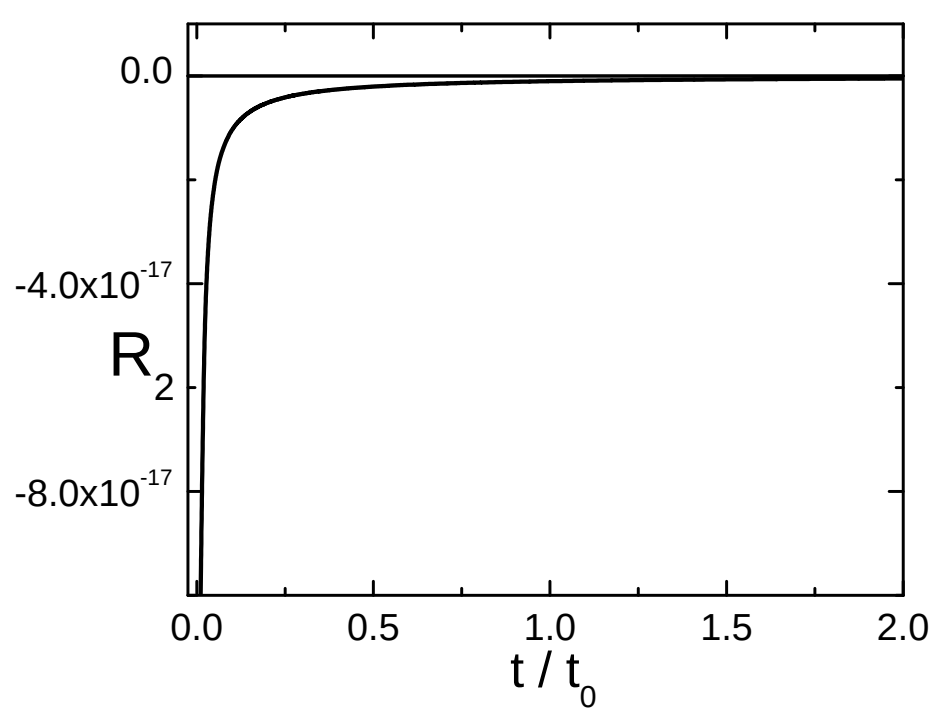

Figure 4: Plot of $R_2$ as a function of time.

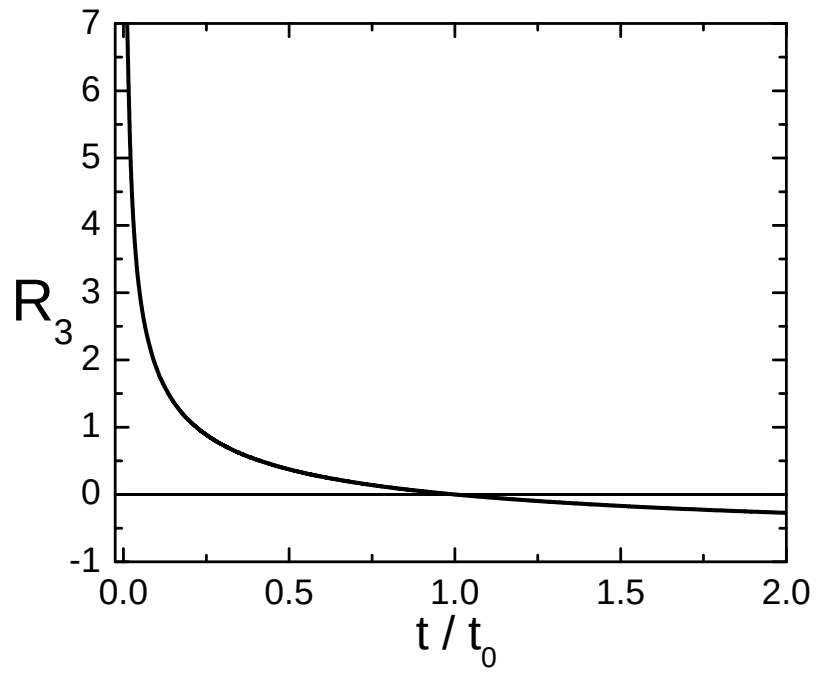

Figure 5: Plot of $R_3$ as a function of time.

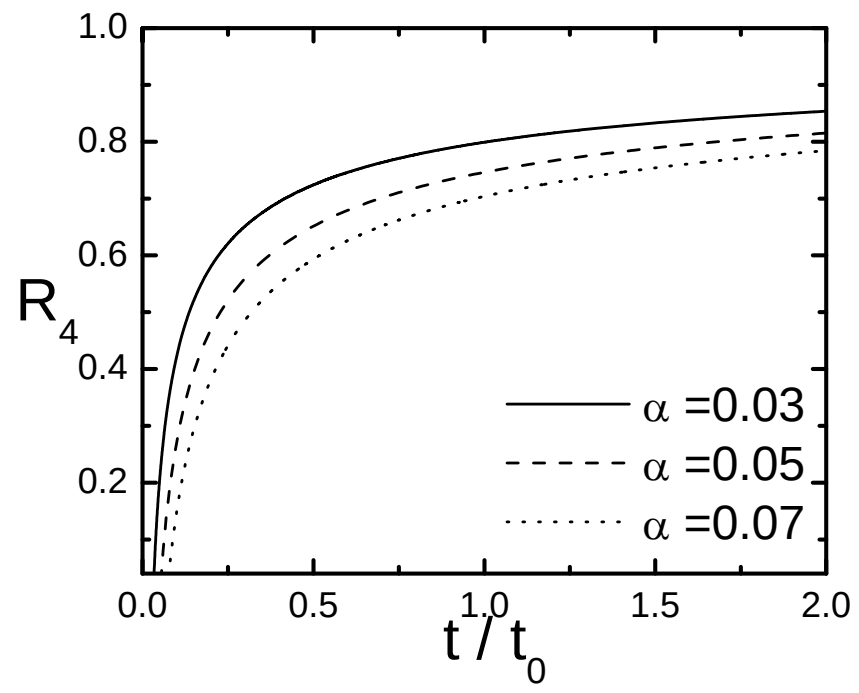

Figure 6: Plot of $R_4$ as a function of time.

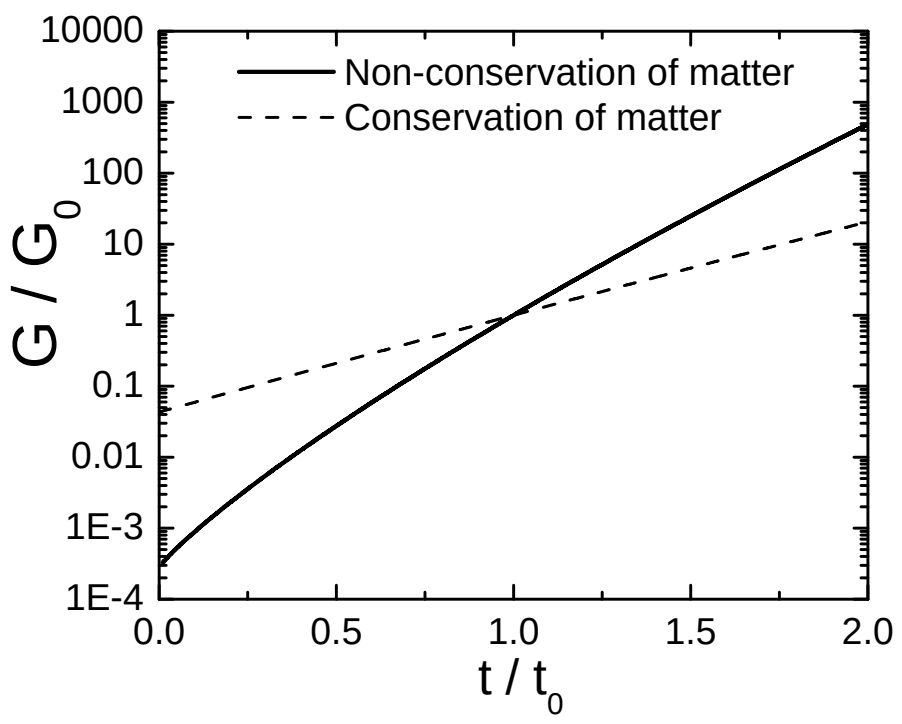

Figure 7: Plot of $G/G_0$ as a function of time.

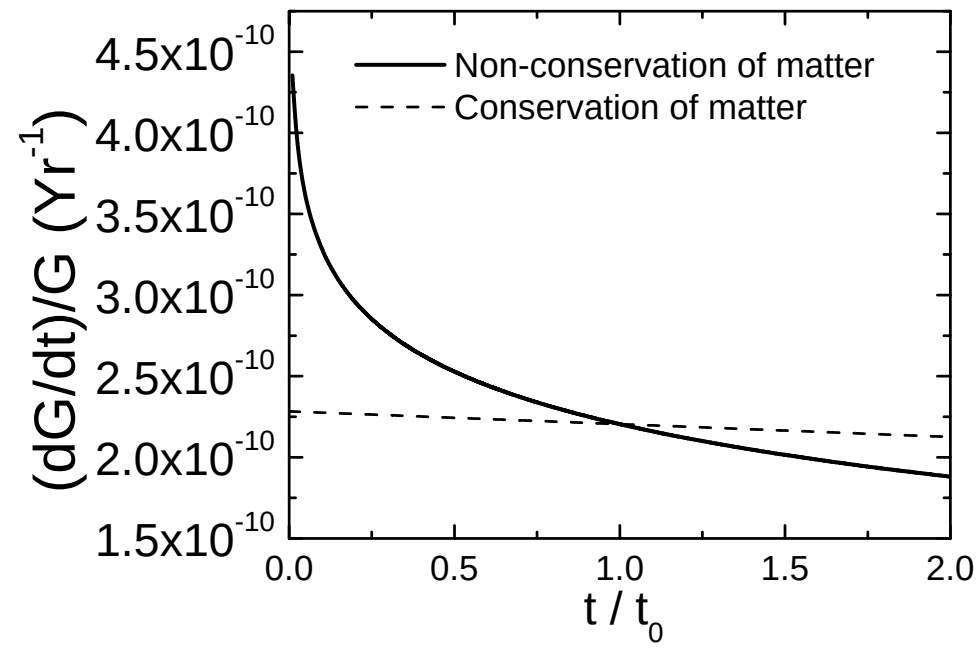

Figure 8: Plot of $\dot{G}/G$ per year as a function of time.